\documentclass[%
 reprint,
superscriptaddress,
 amsmath,amssymb,
]{revtex4-2}

\usepackage{graphicx}
\usepackage{dcolumn}
\newcommand\mathplus{+}
\usepackage{bm}
\usepackage{multirow}
\usepackage{color}
\usepackage{natbib}

\begin{document}

 \preprint{APS/123-QED}

\title{Effect of isovector scalar meson on equation of state of dense matter within  relativistic mean field model}
\author{Virender Thakur}
 \email{virenthakur2154@gmail.com}
 \affiliation{Department of Physics, Himachal Pradesh University, Shimla-171005, India}
 \author{Raj Kumar}
  \email{raj.phy@gmail.com}%
\affiliation{Department of Physics, Himachal Pradesh University, Shimla-171005, India}
\author{Pankaj Kumar}
 \affiliation{Department of Applied Sciences, CGC College of Engineering, Landran, Mohali 140307, India}
\author{Vikesh Kumar}
\affiliation{Department of Physics, Himachal Pradesh University, Shimla-171005, India}
\author{Mukul Kumar}
\affiliation{Department of Physics, Himachal Pradesh University, Shimla-171005, India}
\author{C. Mondal}
\affiliation{Laboratoire de Physique Corpusculaire, CNRS, ENSICAEN, UMR6534, Université de Caen Normandie,
F-14000, Caen Cedex, France}
\author{B.K. Agrawal}
 \email{sinp.bijay@gmail.com}
\affiliation{Saha Institute of Nuclear Physics, 1/AF Bidhannagar, Kolkata 700064, India}
\author{Shashi K. Dhiman}
 \email{shashi.dhiman@gmail.com}
\affiliation{Department of Physics, Himachal Pradesh University, Shimla-171005, India}
\affiliation{School of Applied Sciences, Himachal Pradesh Technical University, Hamirpur-177001, India}

\begin{abstract}
The effects of the isovector-scalar $\delta$-meson field on the properties of   finite nuclei,  infinite nuclear matter
and neutron stars are  investigated within the  Relativistic Mean Field  (RMF) model which includes  non-linear couplings.   Several  parameter sets   (SRV's) are generated to asses the influence of
$\delta$-meson on the properties of neutron star.  These parametrizations
correspond to different values of coupling constant of $\delta$-meson
to the nucleons with remaining ones  calibrated to yield finite nuclei
and infinite nuclear matter  properties consistent with the available
experimental data. It is observed that to fit the properties of finite nuclei and infinite
nuclear matter, a stronger coupling between isovector-vector $\rho$ meson
and nucleons is required in the presence of $\delta$ field.  Furthermore,
the $\delta$-meson is found to affect the radius of canonical neutron
star significantly.  The  value of dimensionless tidal deformability,
${\Lambda}$ for the canonical neutron star also satisfies the constraints
from the waveform models analysis of GW170817 binary neutron star merger
event.  A covariance analysis is performed to estimate  the statistical
uncertainties of the model parameters as well as  correlations among
the model parameters and different observables of interest.
\end{abstract}
\maketitle


\section{Introduction}
Neutron stars are the  densest  objects in the observable universe and deep
knowledge  of the Equation of State (EoS) of the dense matter in beta-equilibrium
is thus required to understand their behavior. It has been shown that the  dense matter EoS  must be treated
relativistically \cite{Walecka1974,Glendenning1991}. For this reason, relativistic mean
field (RMF) models have been widely used to obtain a realistic description of the
properties of finite nuclei, bulk nuclear matter or the properties of neutron stars.
Currently, many different variants  of RMF models with various couplings are in  use to study
the finite  nuclei and neutron star properties \cite{Dutra2014, Oertel2017, Roca-Maza2018}.
Accurate constraints are necessary to understand the limits of these different
types of models. During the last decade, a wide range of astrophysical observations
such as the precise measurement of massive millisecond pulsars using Shapiro delay
technique \cite{Demorest2010, Antoniadis2013}, detection of gravitational wave
generated by binary neutron stars in the GW170817 event by the LIGO-Virgo collaboration
\cite{Abbott2018,Abbott2019}, or the joint mass radius measurement of neutron stars
using X-ray timing technique by NICER collaboration \cite{Miller2019, Riley2019,
Miller2021, Riley2021} started to provide unprecedented new constraints on the dense
matter EoS. They have triggered  plethora of theoretical studies to look at the
dense matter EoS from very different perspectives c.f. Ref. \cite{Lattimer2021} and
references therein. First of its kind
model independent measurement of neutron skin thickness $\Delta r_{np}$ of $^{208}$Pb
\cite{Adhikari2021} and $^{48}$Ca \cite{Adhikari2022} in the Jefferesen lab also
inspired theoretical studies to take a fresher look at the isovector channel of the nuclear
interaction \cite{Essick2021, Reinhard2021, Mondal2022b, Reinhard2022, Yuksel2022}.

Effective mass of nucleon quantifies the momentum dependence of nuclear
force in the medium. It can be quoted for infinite nuclear matter at the Fermi
surface. It is, however, necessary to realize that the concept of effective mass
is different in non-relativistic \cite{Serot1986,Chen2007} and relativistic
formalism \cite{Jaminon1989}.
Nevertheless, it plays some crucial roles to determine various finite nuclear
properties \textit {e.g.} isoscalar giant quadrupole resonance (ISGQR) \cite{Roca-Maza2013},
nucleon nucleon scattering in optical potentials \cite{Li2015b} or even in realizing
various properties of nuclear matter and neutron stars \cite{Malik2018a, Malik2018b}. Recently,
a systematic study was performed using RMF models which assessed
the impact of relativistic (Dirac) effective mass ($M^*$) on the properties of
neutron star \cite{Ghosh2021}. The isovector splitting of the effective
mass, which measures the difference between neutron ($M_n^*$) and proton
($M_p^*$) effective mass, can influence greatly as well the physical properties
of finite nuclei  such as locating the drip-lines  \cite{Woods1997} or
nucleon-nucleus scattering of asymmetric systems \cite{Li2015b}. Its impact increases
manifold in high density environment which can alter thermal and transport
properties of asymmetric matter \cite{Li2004, Xu2015} or neutrino opacities of neutron
star matter \cite{Baldo2014}. To settle its value, even at saturation, remains
a persisting challenge both theoretically \cite{Zuo2005, Dalen2005, Bandyopadhyay1990,
Mondal2017, Mondal2018} and experimentally \cite{ Zhang2016, Li2015b, Kong2017}.

Appearance of isovector splitting of effective mass to the leading order in
RMF models occurs through the isovector-scalar $\delta$ mesons. It impacts the
proton fraction in neutron stars and hence the cooling process of neutron
stars after formation \cite{Roca-Maza2011,Wang2014}. It can also influence the
global properties of neutron stars \cite{Dexheimer2015, Kumar2018, Kumar2021,Malik2018a,Malik2018b,Fan-Li2022}.
A systematic study of RMF models with added freedom in the
isospin channel through $\delta$ meson,
optimized using well constrained finite nuclear properties and extrapolating
at high density to understand the properties of neutron stars and in general
dense matter EoS, can enhance our knowledge on the density dependence of the
isovector channel of nuclear interaction.

The present study is aimed towards  investigating the effects of $\delta$ meson on the dense matter EoS within the framework of RMF model. We generate several parameter sets by varying the coupling strength of
$\delta$ meson to the nucleons such that  the  low density behaviour of
the EOSs remain consistent with the  available finite nuclei data and
a few empirical properties of infinite nuclear matter evaluated at the
saturation density. The properties of neutron stars obtained with these
EOSs are them compared to asses the role  of $\delta$ mesons.

The paper is  organized as follows. In section \ref{model}, the theoretical
framework which is used to construct the EoS for neutron stars  is discussed.
We also discuss the procedure to optimize the coupling constants  and the
method to perform a covariance analysis in the same section. In section \ref{results},
we present our results. We summarize and draw our conclusions in section \ref{conclusion}.

\section{Formalism}\label{model}
\subsection{Theoretical model}
The Lagrangian  density  for  the RMF model used in the present study based on
different non-linear, self and inter-couplings among isoscalar-scalar $\sigma$,
isoscalar-vector $\omega_{\mu}$, isovector-scalar $\delta$ and isovector-vector
$\rho_{\mu}$ meson fields and nucleonic Dirac field $\Psi$ \cite{Dhiman2007,Raj2006},
is given by
\begin{eqnarray}
\label{eq:lbm}
	{\cal L} &=& \sum_q \overline{\Psi}[i\gamma^{\mu}\partial_{\mu}-
(M-g_{\sigma } \sigma - g_{\delta}\delta\cdot\tau)\nonumber\\
	&-&(g_{\omega }\gamma^{\mu} \omega_{\mu}+
\frac{1}{2}g_{\mathbf{\rho}}\gamma^{\mu}\tau\cdot\mathbf{\rho}_{\mu})]\Psi
	+ \frac{1}{2}(\partial_{\mu}\sigma\partial^{\mu}\sigma-m_{\sigma}^2\sigma^2) \nonumber\\
	&-&\frac{\overline{\kappa}}{3!}
g_{\sigma }^3\sigma^3-\frac{\overline{\lambda}}{4!}g_{\sigma }^4\sigma^4
	- \frac{1}{4}\omega_{\mu\nu}\omega^{\mu\nu}+ \frac{1}{2}m_{\omega}^2\omega_{\mu}\omega^{\mu}\nonumber\\
	&+&\frac{1}{4!}\zeta g_{\omega }^{4}(\omega_{\mu}\omega^{\mu})^{2}-\frac{1}{4}\mathbf{\rho}_{\mu\nu}\mathbf{\rho}^{\mu\nu}
	+\frac{1}{2}m_{\rho}^2\mathbf{\rho}_{\mu}\mathbf{\rho}^{\mu}\nonumber\\
	&+&\frac{1}{2}(\partial_{\mu}\delta\partial^{\mu}\delta-m_{\delta}^2\delta^2)
	+\frac{1}{2}c_{1}g_{\omega }^{2}g_{\rho }^2\omega_{\mu}\omega^{\mu}\rho_{\mu}\rho^{\mu}.
\end{eqnarray}
The Dirac effective mass for the nucleons (q) appearing in the Lagrangian density
above is specified as
\begin{equation}
\label{eq:nucleon}
	M^{*}_{q} =( M - g_{\sigma}\sigma - g_{\delta}\delta\cdot \tau),
\end{equation}
where, $\tau =1 (-1)$ for $q=$ neutron (proton). Following the Euler-Lagrange
formalism one can readily find the expressions for energy density ${\cal E}$ and
pressure $P$ as a function of density from Eq. (\ref{eq:lbm}) \cite{Glendenning2000}.

\subsection{Optimization and covariance analysis}
In the present study, five new relativistic interactions SRV00, SRV01,
SRV02, SRV03, and SRV04  have been generated for the Lagrangian density
given by Eq. (\ref{eq:lbm}) to investigate the effect of $\delta$ meson
on the properties of finite nuclei and neutron star matter.  Here,
SRV00, SRV01, SRV02, SRV03 and SRV04 parametrizations correspond to
different value of the coupling of $\delta$-meson to the nucleon i.e.
$g_{\delta}$ = 0.0, 1.0, 2.0, 3.0 and 4.0 respectively.  As the effect of
$\delta$ meson is predominantly important at suprasaturation densities,
one can \textit{a-priori} anticipate its insignificant  impact in finite nuclei,
which is primarily  sensitive to the EoS at subsaturation densities. This
is the reason why we kept fixed the $g_{\delta}$ at aforementioned values
optimizing the rest of the parameters in Eq. (\ref{eq:lbm}). This is not
far from the strategy recently used by Li \textit{et. al.} in Ref. \cite{Fan-Li2022}.
The parameters of the model are obtained  by fitting the experimental data
\cite{Wang2021} on binding energies ($BE$) and charge rms radii ($r_{ch}$) \cite{Angeli2013}
of some spherical nuclei $^{16,24}$O, $^{40,48}$Ca, $^{56,78}$Ni,
$^{88}$Sr,$^{90}$Zr, $^{100,116,132}$Sn and $^{208}$Pb. For the open shell
nuclei, the pairing has been included using BCS formalism  with constant
pairing gaps \cite{Ring1980,Karatzikos2010} that are taken from the 
nucleon  separation energies  of neighboring nuclei \cite{Wang2021}. Neutron and proton pairing gaps  are evaluated by using  fourth order finite difference mass formula (five point difference) \cite{Duguet2001}. The neutron and proton pairing gaps ($\Delta_{n}$,$\Delta_{p}$)  in  MeV for the open shell nuclei  are  $^{88}Sr$(0.0,1.284), $^{90}Zr$(0.0,1.239)  and $^{116}Sn$(1.189,0.0).  The neutron pairing gap for $^{24}O$ practically vanishes since the first unoccupied orbit 1$d_{3/2}$ is almost 4.5 MeV above the completely filled 2$s_{1/2}$ orbit \cite{Chen2015,Mondal2016}. The pairing correlation energies for a fix gap $\Delta$ is calculated by using the paring window of 2$\hbar\omega$, where $\hbar\omega$ = $45 A^{-1/3}$ - $25 A^{-2/3}$ MeV \cite{Raj2006}.
We also incorporated the recently measured neutron skin thickness of $^{208}$Pb
using the parity violating electron scattering experiment \cite{Adhikari2021}
in our fit data.

\begin{table*}[t]
\centering
\caption{\label{tab:table1} 
SRV parameter  sets for  the Lagrangian of RMF model as given
in Eq.(\ref{eq:lbm}).
The parameter $\overline{\kappa}$ is in  fm$^{-1}$. The values of meson masses  $m_{\sigma}$,
$m_{\omega}$, $m_{\rho}$ and $m_{\delta}$  are in  MeV. The nucleonic mass (M) and
meson masses  ($m_{\omega}$, $m_{\rho}$ and $m_{\delta}$) are  taken as 939, 782.5, 762.468
and 980 MeV, respectively.  The values of $\overline{\kappa}$, $\overline{\lambda}$,
and ${c_{1}}$ are multiplied by $10^{2}$.} 
\begin{tabular}{ccccccc}
\hline
\hline
\multicolumn{1}{c}{${\bf {Parameters}}$}&
\multicolumn{1}{c}{{\bf SRV00}}&
\multicolumn{1}{c}{{\bf SRV01}}&
\multicolumn{1}{c}{{\bf SRV02}}&
\multicolumn{1}{c}{{\bf SRV03}}&
\multicolumn{1}{c}{{\bf SRV04}}\\
\hline
${\bf g_{\sigma}}$  &10.3109$\pm$0.1109&10.3442$\pm$0.0978&10.3344$\pm$0.1135&10.3723$\pm$0.0999&10.3735$\pm$0.1294\\
${\bf g_{\omega}}$  &13.1772$\pm$0.1621 & 13.2508$\pm$0.1379&13.2137$\pm$0.1414&13.3113$\pm$0.1269&13.2984$\pm$0.1815\\
${\bf g_{\rho}}$ &10.8834$\pm$1.1730&11.2832$\pm$1.0110 &11.5834$\pm$0.9676 &12.4834$\pm$1.1274&13.1833$\pm$0.8043\\
${\bf \overline {\kappa}}$&1.8509$\pm$0.0500 &1.8624$\pm$0.0700 &1.8780$\pm$0.0200 &1.8242$\pm$0.0500&1.8490$\pm$0.0800\\
${\bf \overline {\lambda}}$&-0.05151$\pm$0.0700&-0.05803$\pm$0.0600&-0.06060$\pm$0.0600& -0.04934$\pm$0.0800&-0.05621$\pm$0.0800\\
${\bf{\zeta}}$&0.02116$\pm$0.0017&0.02050$\pm$0.0013&0.02017$\pm$0.0011& 0.02177$\pm$0.0012&0.02089$\pm$0.0027\\
${\bf c_1}$ &4.60038$\pm$2.3600&4.49219$\pm$1.9200&4.20853$\pm$1.5600 &3.66169$\pm$1.4900 &3.01068$\pm$0.8800\\
${\bf m_{\sigma}}$ &501.9596$\pm$0.9230&501.0200$\pm$1.0071 &501.6638 $\pm$1.3141&500.7480$\pm$1.2629&501.0215$\pm$1.3663\\
\hline
\hline
\end{tabular}
\end{table*}
The optimization of the parameters ($\textbf{p}$) appearing in the Lagrangian (Eq. \ref{eq:lbm})
is done by using the simulated annealing method (SAM) \cite{Agrawal2005, Burvenich2004,
Kirkpatrick1984} by following $\chi^{2}$ minimization procedure  which is given
as,
\begin{equation}
	{\chi^2}(\textbf{p}) =  \frac{1}{N_d - N_p}\sum_{i=1}^{N_d}
\left (\frac{ M_i^{exp} - M_i^{th}}{\sigma_i}\right )^2 \label {chi2},
\end{equation}
where $N_d$ is the number of  experimental data points and $N_p$ is the number
of fitted  parameters. The $\sigma_i$ denotes adopted errors \cite{Dobaczewski2014,Mondal2015}
and $M_i^{exp}$ and $M_i^{th}$ are the experimental and the corresponding
theoretical values, respectively, for a given observable.  The minimum value of
${{\chi}}^{2}_{0}$  corresponds to the optimal values $\textbf{p}_{0}$ of the parameters.

Once the optimized parameter set is obtained, the correlation coefficient
between two quantities Y and Z, can be calculated by covariance analysis
\cite{Brandt1997,Reinhard2010,Fattoyev2011,Dobaczewski2014,Mondal2015} as
\begin{equation}
\label{covariance}
    \textit{r}_{\scriptstyle{YZ} }= \frac{\overline{\Delta{Y} \Delta{Z} }}{\sqrt{\overline{\Delta{Y^2}}
  \quad \overline{\Delta{Z^2}}}} ,
\end{equation}
where covariance between Y and Z is expressed as
\begin{equation}
\label{error}
    \overline{\Delta{Y}\Delta{Z}} = \sum_{\alpha\beta} \left( \frac{\partial{Y}}
{\partial{p}_{\alpha}}\right) _{\textbf{p}_0} C_{\alpha\beta}^{-1}
\left( \frac{\partial{Z}}{\partial{p}_{\beta}}\right) _{\textbf{p}_0}.
\end{equation}
Here, $C_{\alpha\beta}^{-1}$ is an element of inverted curvature matrix given by
\begin{equation}
\label{matrix}
    \textit{C}_{\alpha\beta} = \frac{1}{2}\left(\frac{\partial^2
	\chi^2(\textbf{p})}{\partial{p}_{\alpha}\partial{p}_{\beta}}\right)_{\textbf{p}_{0}}.
\end{equation}
The standard deviation, $\overline{\Delta{Y}^2}$, in Y can be computed using
Eq. (\ref{error}) by substituting Z = Y.

\section{Results and Discussion}\label{results}
We have obtained five different parameter sets corresponding to different values of $g_\delta$  by calibrating the remaining prameters to a suitable set of  on finite nuclei as described  earlier. 
 All parametrizations
obtained in the present work give equally good fit to the properties of finite
nuclei which were used for the optimization procedure. In  Table \ref{tab:table1}  we  display   optimum values of the  model parameters for all the five SRV parameter sets along with the 
uncertainties  on them  computed using  Eq. (\ref{error}). It can be seen   that
the parameter $g_{\rho}$ increases with the increase in value of $g_{\delta}$. A
larger value of $g_{\rho}$ is required in the presence of the $\delta$-field to
fit the properties of finite nuclei. As the contribution of the $\delta$-field
is attractive, increased binding due to the $\delta$-field has to be compensated  by
the higher value of the repulsion by the $\rho$-field. The parameter $g_{\rho}$
has its lowest value for  SRV00  parametrization  ($g_{\delta}$=0). For any
finite value of $\delta$-coupling ($g_{\delta}>0$) i.e. for SRV01, SRV02, SRV03
and SRV04 parametrizations, the strength of $\rho$-coupling ($g_{\rho}$) increases
gradually. The cross-coupling between the $\omega_{\mu}$ and $\rho_{\mu}$ fields
quantified by the term $c_{1}$ decreases slightly from 4.6003 to 3.0107 as the
value of coupling constant $g_{\delta}$ increases from 0.0 to 4.0 corresponding to
different SRV parametrizations.

\begin{table*}
\centering
\caption{\label{tab:table2}
The values of binding energy (BE) and charge radii ($r_{ch}$) of fitted nuclei along with theoretical errors
obtained for different SRV parametrizations. The  corresponding experimental values
\cite{Wang2021,Angeli2013} are also listed. The value of neutron skin thickness ($\Delta r_{np}$) for $^{208}$Pb
is  given along with the experimental data \cite{Adhikari2021}. The adopted errors on the observables  ($\sigma$) used for optimisation of parameters are also displayed. The value of BE
are given in units of MeV and  $r_{ch}$ and $\Delta r_{np}$ are in  fm.  } 
\begin{tabular}{|c|c|c|c|c|c|c|c|c|}
\hline
\hline
\multicolumn{1}{|c|}{${\bf {Nucleus}}$}&
\multicolumn{1}{|c|}{${\bf {Observables}}$}&
\multicolumn{1}{|c|}{{\bf Exp.}}&
\multicolumn{1}{|c|}{{\bf $\sigma$}}&
\multicolumn{1}{|c|}{{\bf SRV00}}&
\multicolumn{1}{|c|}{{\bf SRV01}}&
\multicolumn{1}{|c|}{{\bf SRV02}}&
\multicolumn{1}{|c|}{{\bf SRV03}}&
\multicolumn{1}{|c|}{{\bf SRV04}}
\\
\hline
$^{16}O$ &BE&127.62&4.0&128.67$\pm$0.52&128.83$\pm$0.50&128.94$\pm$0.55&128.90$\pm$0.51&128.94$\pm$0.61\\
 &$r_{ch}$&2.699&0.04&2.709$\pm$0.008&2.712$\pm$0.023&2.711$\pm$0.031&2.710$\pm$0.013&2.710$\pm$0.023\\
 \hline
 $^{24}O$ &BE&168.96&1.0&169.95$\pm$0.96 &169.38$\pm$0.90&169.64$\pm$0.89&168.90$\pm$1.01&169.00$\pm$0.93\\
 \hline
 $^{40}Ca$&BE &342.04&3.0&343.40$\pm$0.69&343.14$\pm$0.65&343.18$\pm$0.53&344.41$\pm$0.63&344.39$\pm$0.91\\
 &$r_{ch}$&3.478&0.02&3.454$\pm$0.014 &3.457$\pm$0.007&3.455$\pm$0.016&3.455$\pm$0.014&3.455$\pm$0.017\\
 \hline
 $^{48}Ca$ &BE&415.97&1.0&415.59$\pm$0.54&415.15$\pm$0.52&415.37$\pm$0.45&415.37$\pm$0.63&415.48$\pm$0.52\\
 &$r_{ch}$&3.477&0.02&3.468$\pm$0.016&3.469$\pm$0.014&3.468$\pm$0.014&3.467$\pm$0.014&3.467$\pm$0.022\\
 \hline
 $^{56}Ni$ &BE&484.01&5.0&482.05$\pm$1.28&482.39$\pm$1.21&482.28$\pm$1.31&483.12$\pm$1.48&483.22$\pm$1.56\\
 &$r_{ch}$&3.750&0.02&3.712$\pm$0.022&3.709$\pm$0.017&3.708$\pm$0.14&3.707$\pm$0.017&3.705$\pm$0.013\\
 \hline
 $^{78}Ni$&BE&642.56 &2.0&641.39$\pm$1.03 &640.01$\pm$0.97&641.05$\pm$1.01&640.21$\pm$1.08&640.18$\pm$1.07\\
 \hline
 $^{88}Sr$ &BE&768.42&1.0&768.24$\pm$0.60&767.72$\pm$0.57&767.84$\pm$0.58&768.39$\pm$0.60&768.40$\pm$0.62\\
 &$r_{ch}$&4.219&0.02&4.226$\pm$0.016&4.227$\pm$0.014&4.227$\pm$0.013&4.225$\pm$0.021&4.225$\pm$0.0146\\
 \hline
 $^{90}Zr$ &BE&783.81&1.0&783.92$\pm$0.69 &783.58$\pm$0.64&783.59$\pm$0.68&784.37$\pm$0.69&784.36$\pm$0.68\\
 &$r_{ch}$&4.269&0.02&4.280$\pm$0.019&4.280$\pm$0.019&4.280$\pm$0.013&4.278$\pm$0.029&4.278$\pm$0.015\\
 \hline
 $^{100}Sn$&BE&825.10&2.0&826.60$\pm$1.28&826.95$\pm$1.24&827.56$\pm$1.14&827.09$\pm$1.23&827.01$\pm$1.66\\
 \hline
 $^{116}Sn$ &BE&988.67&2.0&988.34$\pm$0.83&987.74$\pm$0.73&987.72$\pm$2.18&988.49$\pm$0.82&988.25$\pm$0.88\\
 &$r_{ch}$&4.627&0.02&4.617$\pm$0.016 &4.618$\pm$0.014&4.617$\pm$0.009&4.615$\pm$0.012&4.615$\pm$0.020\\
 \hline
 $^{132}Sn$&BE&1100.22 &1.0&1101.38$\pm$0.84&1100.58$\pm$0.80&1101.48$\pm$0.77&1100.59$\pm$0.86&1100.70$\pm$0.82\\
 &$r_{ch}$&4.709&0.02&4.721$\pm$0.019&4.721$\pm$0.019&4.720$\pm$0.010&4.719$\pm$0.012&4.717$\pm$0.011\\
 \hline
 $^{208}Pb$ &BE&1636.34&1.0&1636.58$\pm$1.03&1635.98$\pm$0.98&1636.32$\pm$0.96&1636.27$\pm$1.05&1636.10$\pm$1.02\\
 &$r_{ch}$&5.501&0.02&5.530$\pm$0.015 &5.531$\pm$0.014&5.529$\pm$0.014&5.527$\pm$0.021&5.526$\pm$0.011\\
 &$\Delta r_{np}$&$0.283\pm0.071$&0.071&0.222$\pm$0.032&0.223$\pm$0.028&0.217$\pm$0.026&0.215$\pm$0.035&0.214$\pm$0.029\\
\hline
\hline
\end{tabular}
\end{table*}
\begin{figure}
\centering
\includegraphics[trim=0 0 0 0,clip,scale=0.6]{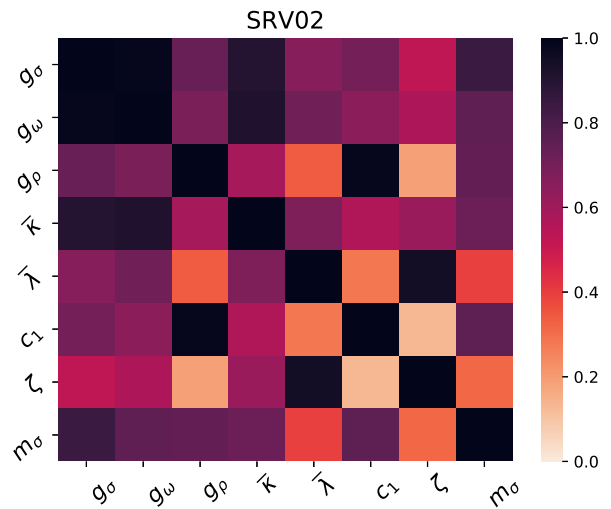}
\caption{\label{parameter_corr} (Color online) Correlation coefficients
(absolute values) among the model parameters for SRV02 parametrization
for  the Lagrangian given by Eq. (\ref{eq:lbm}).}
\end{figure}
In Table \ref{tab:table2}, different observables  fitted in the present
work, their experimental values \cite{Wang2021,Angeli2013}, adopted errors $\sigma$  on them
 \cite{Klupfel2009} along with the calculated values for   different SRV
parametrizations are displayed. The  estimated uncertanties are also listed   for the fitted observables.  The fitted values of finite nuclei  properties
are quite close to their experimental counterparts. The root mean
square (rms) errors  on the $BE$  are found to be in the range $ 1.50 - 1.86$ MeV,
and the ones for $r_{ch}$ are found to be  $~ 0.02 $ fm  for the different parameterizatons. It is quite interesting to observe that even though $g_{\delta}$ influences the coupling
$g_{\rho}$, the isovector sensitive observable $\Delta r_{np}$ varies only slightly
$\sim 0.01$ fm across the different SRV models obtained in the present work. This
observation is quite similar to the one obtained by Li \textit{et. al.}
\cite{Fan-Li2022}.

\begin{table*}
\centering
\caption{\label{tab:table3}
The bulk nuclear matter properties at saturation density for SRV parametrizations are listed:
$\rho_{0}$, E/A, K, J, L and $M^{*}/M$ denotes the saturation density, binding energy
per nucleon, incompressibility coefficient, symmetry energy, the slope of symmetry
energy, and the ratio of effective nucleon mass to the nucleon mass, respectively.}
\begin{tabular}{cccccc}
\hline
\hline
\multicolumn{1}{c}{${\bf {Parameters}}$}&
\multicolumn{1}{c}{{\bf SRV00}}&
\multicolumn{1}{c}{{\bf SRV01}}&
\multicolumn{1}{c}{{\bf SRV02}}&
\multicolumn{1}{c}{{\bf SRV03}}&
\multicolumn{1}{c}{{\bf SRV04}}\\
\hline
${\bf{\rho_{0} ~(fm^{-3})}}$&0.149$\pm$0.003&0.149$\pm$0.002 &0.149$\pm$0.002&0.149$\pm$0.003&0.149$\pm$0.008\\
${\bf E/A ~(MeV)}$ &-16.11$\pm$0.06 & -16.11$\pm$0.05&-16.09$\pm$0.05&-16.12$\pm$0.06&-16.11$\pm$0.04\\
${\bf K ~(MeV)}$ &223.94$\pm$8.57&221.78$\pm$9.95 &222.05$\pm$5.47&221.72$\pm$10.62&221.11$\pm$23.20\\
${\bf J ~(MeV)}$&33.49$\pm$1.82&33.75$\pm$1.77&33.31$\pm$1.78&33.54$\pm$2.13&33.34$\pm$2.08\\
${\bf L ~(MeV)}$&65.23$\pm$15.37&63.82$\pm$13.50&61.49$\pm$13.22&58.06$\pm$15.93&55.31$\pm$13.76\\
${\bf ~M^{*}/M}$&0.606$\pm$0.013&0.0.602$\pm$0.010&0.603$\pm$0.005& 0.601$\pm$0.009&0.600$\pm$0.009\\
\hline
\hline
\end{tabular}
\end{table*}
In Table \ref{tab:table3}, we present the results for the properties of
Symmetric Nuclear Matter (SNM) such as binding energy per nucleon (E/A),
incompressibility (K), the ratio of effective mass to the mass of nucleon
($M^{*}/M$) along with symmetry energy coefficient (J),  and its slope  (L),  all are  evaluated 
at the saturation density ($\rho_{0}$). We also quote the theoretically
calculated error on them. The results  are presented for all five SRV
parametrizations. The value of E/A lies in the range $-16.09$ to $-16.12$
for the five parametrizations. The  value of J and L obtained by our
parametrizations are  consistent with the constraints from observational
analysis J = 31.6 $\pm$ 2.66 MeV and L = 58.9 $\pm$ 16 MeV \citep{Li2013,
Yue2022}. The value of K is also in agreement with the value 240 $\pm$ 20
MeV determined from isoscalar giant monopole resonance (ISGMR) for  $^{90}$Zr
and $^{208}$Pb nuclei \citep{Colo2014,Piekarewicz2014}. It can also be seen
from Table \ref{tab:table3} that  the mean values of the slope of
of  symmetry energy (L) for SRV parametrizations decrease with the increase
in the value of $g_{\delta}$. The average value of L decreases from 65.23 MeV
in SRV00 to 55.31 MeV for SRV04. It can be noted that the isoscalar properties
(E/A, K, $\rho_{0}$ and $M^{*}/M$) are well constrained  for all SRV
parametrizations. The only exception is the error on K in case of SRV04,
where the error is almost 10\% of its central value. But in the
isovector sector, the percentage error on
the slope of symmetry energy (L) are consistently on the larger side for all
SRV parametrizations.

A great deal of importance to perform covariance analysis in theoretical studies
has been pointed out recently \cite{Reinhard2010, Dobaczewski2014}. It not only enables one to quote statistical uncertainties on model parameters or any calculated
observables, but also privedes complementary information about the
sensitivity of the parameters to physical observables, redundancies among fitted
observables or interdependences among model parameters. As our primary objective is
not to establish an ultimate model, for demonstrative purpose, we will discuss the
results of covariance analysis as outlined in Section \ref{model}B, only for the
model SRV02. The results  for other parameter sets are quite similar (not shown here).
In Fig. \ref{parameter_corr}, the correlation coefficients between different
model parameters appearing in Eq. (\ref{eq:lbm}) are outlined for SRV02
parametrization. A strong correlation is found between the several pairs
of model parameters,  like, $g_{\sigma}$ and $g_{\omega} $, $ g_{\rho}$ and $c_{1}$ and
$\overline{\lambda}$ and $\zeta$ with correlation coefficients 0.99, 0.98 and
0.95, respectively. These interdependences mean that if one of these pairs are fixed at a
particular value, the other must attain the precise value as suggested by their
correlation to satisfactorily obtain the fit data. The results obtained for the  correlations among
model parameters presented  in
Fig. \ref{parameter_corr} are  quite similar to the obtained  in Refs. \cite{Chen2014, Mondal2022b}. 
Anticipating the strong correlation between L and $\Delta r_{np}$
which is shown later (see Fig. \ref{snm_skin_corr}),
this may be attributed to the large experimental error on the
$\Delta r_{np}$ for $^{208}$Pb, which also led us choosing a rather large
adopted error during optimization. The theoretical errors on the $\Delta r_{np}$
of $^{208}$Pb nucleus are found to be 0.032, 0.028, 0.026, 0.0353 and 0.029 fm
for SRV00, SRV01, SRV02, SRV03, and SRV04 parametrizations, respectively. These
are much smaller compared to the adopted error (0.071 fm, which is also the
experimental error obtained in Ref. \cite{Adhikari2021}).

\begin{figure}
\centering
\includegraphics[trim=0 0 0 0,clip,scale=0.61]{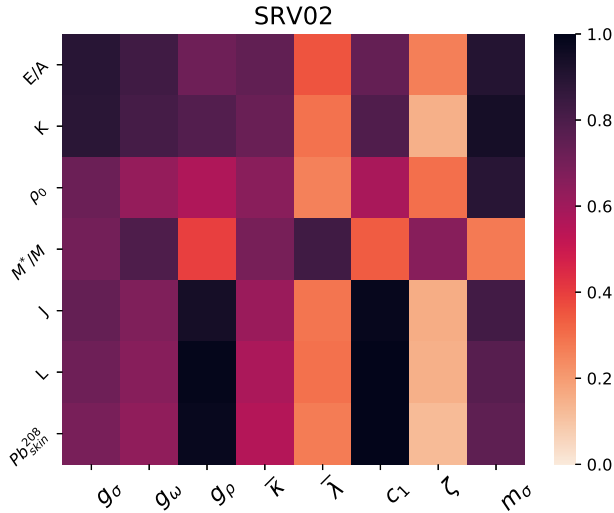}
\caption{\label{par_vs_snm_skin} (Color online) Correlation coefficients
(absolute values) between the model parameters and a set of physical
	observables for SRV02 parametrization (see text for details).}
\end{figure}
We now display in Fig. \ref{par_vs_snm_skin}, the correlation coefficients
between the model parameters apearing in  Lagrangian (Eq. \ref{eq:lbm})  and the
different properties of interest corresponding to the SNM and  the  density dependent symmetry energy  as
displayed in Table \ref{tab:table3} and the $\Delta r_{np}$ of $^{208}$Pb
for SRV02. A  strong correlation is observed between the isovector parameter
$g_{\rho}$ with the symmetry energy coefficient (J), its slope  (L) and
$\Delta r_{np}$ of $^{208}$Pb. The vector mixing parameter $c_{1}$ is also
found to have a strong correlation with the J and L. This strong correlation
is anticipated, as $c_{1}$ and $g_{\rho}$ are strongly correlated to each
other, which was observed in Fig. \ref{parameter_corr}. It is also realized
from Fig. \ref{par_vs_snm_skin} that bulk properties of SNM like E/A, K,
$\rho_{0}$ and $M^{*}/M$ have strong correlations with isoscalar coupling
parameters $g_{\sigma}$, $g_{\omega}$ and $\overline{\kappa}$. This study is
quite consistent with previous calculations in the literature \cite{Chen2014,
Mondal2016, Reinhard2022}. 
\begin{figure}
\centering
\includegraphics[trim=0 0 0 0,clip,scale=0.6]{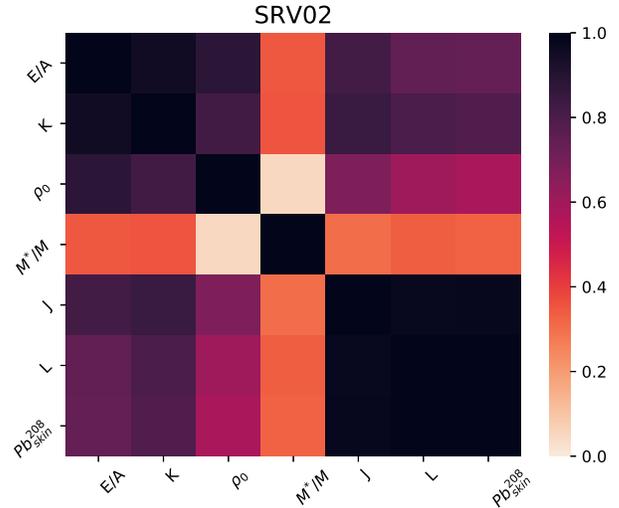}
\caption{\label{snm_skin_corr} (Color online) Correlation coefficients
(absolute values) for few bulk nuclear matter properties and neutron
skin of $^{208}$Pb for SRV02 parametrization.}
\end{figure}
In Fig. \ref{snm_skin_corr}, we display the correlation coefficients among
the different observables in graphical form, particularly which were also studied in Fig.
\ref{par_vs_snm_skin}. In the isoscalar sector the only strong correlation
observed is between binding energy per nucleon (E/A) and incompressibility
coefficient (K). The  K also shows some mild correlations with all other observables
displayed in the figure.  The symmetry energy J and its
slope parameter L are found to be strongly correlated. As mentioned earlier in
the discussion of Table \ref{tab:table3}, we observe a strong correlation of the
neutron skin thickness of $^{208}$Pb with J and L. These results are also in
line with the earlier ones \cite{Chen2014}.

\begin{figure}
\centering
\includegraphics[trim=0 0 0 0,clip,scale=0.40]{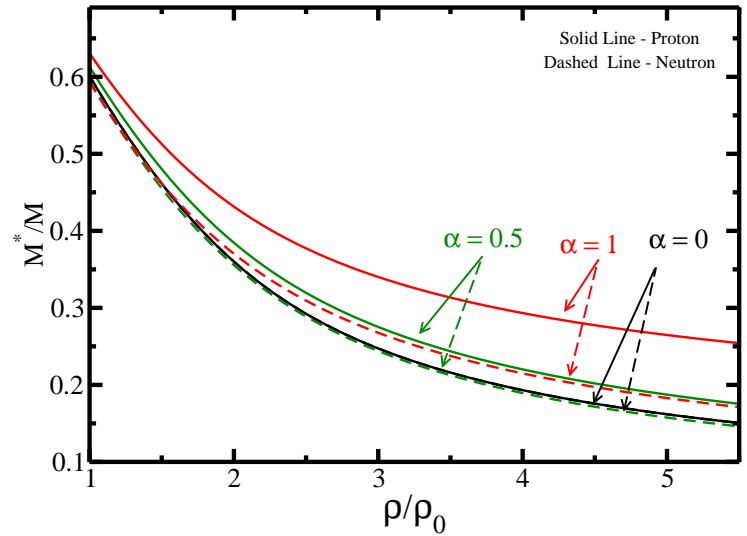}
\caption{\label{fig:fig2} (Color online) Effective masses of proton and neutron
for a few values of asymmetry parameter $\alpha$ for SRV04 parametrization.}
\end{figure}

It is quite important to emphasize 
 that we  kept fixed the strength of the coupling of $\delta$ meson 
to different values and optimized the rest. This might be partially responsible 
to impart a strong correlation among the isovector sensitive parameters 
$g_{\rho}, c_1$ to J, L or $\Delta r_{np}$ of $^{208}$Pb (see Fig. 
\ref{par_vs_snm_skin}) to reproduce the fitted data within bounds. It 
further gets clarified in the strong correlations among $\Delta r_{np}$ 
($^{208}$Pb), J and L in Fig. \ref{snm_skin_corr}. Anticipating the results 
obtained for neutron stars which are discussed later, this strong correlation 
somewhat restricts the behavior of the matter at high densities in the isovector 
channel, resulting in a monotonic increasing in the radius and tidal deformability 
of 1.4 $M_{\odot}$  (see Table \ref{tab:table4}) with the increase of the 
$\delta$ meson coupling $g_{\delta}$. A full optimization is thus needed 
with suitable data in the future including $g_{\delta}$, to understand 
this behavior further. In the present work we have included the simplest form of $\delta$-meson coupling ($g_{\delta}$) to the nucleons. Furthermore, a higher order mixed scalar interactions of $\delta$ meson has a large influence on the symmetry energy and its density dependence. This enables  one to have flexibility to vary  the behaviour of EoS at high density and gives a large influence on the properties of neutron stars \cite{Zabari2019,Zabari2019_2,Miyatsu2022}.
To study  the effects of $\delta$-meson on nucleon mass, in Fig. \ref{fig:fig2},
the effective mass of proton and neutron are plotted as a function of baryon
density for three values of  asymmetry parameter $\alpha$= 0, 0.5, 1
($\alpha$= $\frac{\rho_{n}-\rho_{p}}{\rho_{n}+\rho_{p}}$) for SRV04
parametrization, which has the largest value of $g_{\delta}$ amongst all SRV
variants obtained in the present work.  The asymmetry parameter $\alpha$ = 0.0
represents the SNM and $\alpha$=1 corresponds to the pure neutron matter (PNM).
It is clear from the Eq. (\ref{eq:nucleon})  that the presence of $\delta$-meson
leads to  splitting of nucleon mass.  For SNM, there is no splitting of the
nucleon mass.  In  Fig.
\ref{fig:fig2}, the solid (dashed) lines depict the  effective mass of proton
(neutron) for $\alpha$= 0, 0.5 and 1.0. One can observe from the figure that
the effective proton mass is larger than the neutron effective mass.
The splitting of the proton and neutron effective masses due to the $\delta$-meson
can be important in the highly asymmetric system like a neutron star or
supernova environment. At the center of a neutron star the density can reach
$\sim$ 5-6 $\rho_0$ and $\alpha\sim$ 0.7-0.8. One can readily estimate the amount
of splitting in the effective mass in this situation looking at Fig. \ref{fig:fig2}.
It can also affect the transport properties of neutron star matter \cite{Kubis1997}.

\begin{figure}
\centering
\includegraphics[trim=0 0 0 0,clip,scale=0.56]{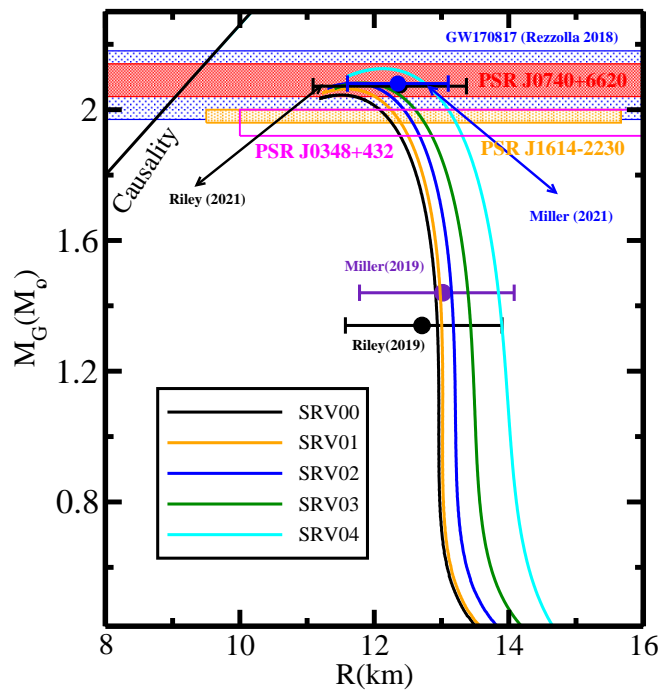}
\caption{\label{mr} (Color online) Mass-Radius relation of neutron star for SRV parametrizations.}
\end{figure}
\begin{table*}
 \caption{\label{tab:table4}The properties of nonrotating  neutron star for the various EOSs computed with  SRV parameter sets are presented along with the theoretical errors on them. M$_G$ and  R$_{max}$  denote the Maximum Gravitational mass and corresponding  radius. The
 values for R$_{1.4}$ and $\Lambda_{1.4}$ denote radius and  dimensionless tidal deformability at 1.4M$_\odot$.
}
 \begin{tabular}{ccccccc}
\hline
\bf{No.} &\bf{EOS}& \bf{$M_G$}& \bf{R$_{max}$ } & \bf{R$_{1.4}$}&\bf{$R_{2.0}$}&${\bf{\Lambda_{1.4}}}$\\
 & &(M$_{\odot}$) & (km)& (km) &(km) & \\
 \hline

1.&SRV00&2.04$\pm$0.03&11.48$\pm$0.08&12.92$\pm$0.22&12.07$\pm$0.31&484.16$\pm$62.10\\
2.&SRV01&2.06$\pm$0.02&11.55$\pm$0.13&12.99$\pm$0.20&12.24$\pm$0.21&504.95$\pm$58.02\\
3.&SRV02&2.07$\pm$0.02&11.67$\pm$0.07&13.16$\pm$0.13&12.43$\pm$0.14&565.52$\pm$60.33\\
4.&SRV03&2.08$\pm$0.02&11.81$\pm$0.11&13.41$\pm$0.21&12.60$\pm$0.16&652.60$\pm$52.01\\
5.&SRV04&2.13$\pm$0.06&12.11$\pm$0.23&13.86$\pm$0.21&13.09$\pm$0.33&783.96$\pm$70.03\\

\hline
\end{tabular}
\end{table*}
To assess the impact of $\delta$-meson on the global properties of neutron star,
we plot the gravitational mass ($M_G$) of non-rotating neutron star as a function of radius
for all SRV parametrizations in Fig. \ref{mr}. The maximum mass ($M_{max}$) and
the corresponding radius ($R_{max}$) of neutron star for all the models obtained here
lie in the range 2.04 - 2.13 $M_{\odot}$ and 11.48 - 12.11 km, respectively. This
satisfies the recently measured radius of PSRJ0740+6620 with $12.45^{+0.65}_{-0.65}$ km
by NICER collaboration \cite{Miller2021,Riley2021}. The radius of neutron star of
2$M_{\odot}$ is also in accordance with the observational data of PSRJ0740+6620
by NICER \cite{Miller2021,Riley2021}. The maximum mass of the neutron star attained
by various SRV parametrizations supports the constraint from PSRJ0740+6620 with the mass
of 2.08 $\pm$ 0.07 $M_{\odot}$ \cite{Cromartie2019,Fonseca2021}. It is observed that
the radius  ($R_{1.4}$) of  neutron star with mass 1.4$M_{\odot}$, can be significantly
affected by the presence of $\delta$-meson as we move from parametrization set SRV00
to SRV04. The value of $R_{1.4}$ with the inclusion of TM1 crust EoS \cite{Sugahara1994}
lies in the range 12.92-13.86 km, which is also in line with the range proposed in
Ref. \cite{Miller2021,Reed2021}. It is observed that the radius $R_{1.4}$ increases by
7.27 \%  and the maximum mass of neutron star changes by 4.4 \% from SRV00 to SRV04
parametrizations with the variation of coupling  $g_{\delta}$= 0.0 to 4.0. This change
in the neutron star properties may be attributed to the impact of $\delta$-meson, which
affects high-density  behavior of asymmetric nuclear matter.

\begin{figure}
\centering
\includegraphics[trim=0 0 0 0,clip,scale=0.42]{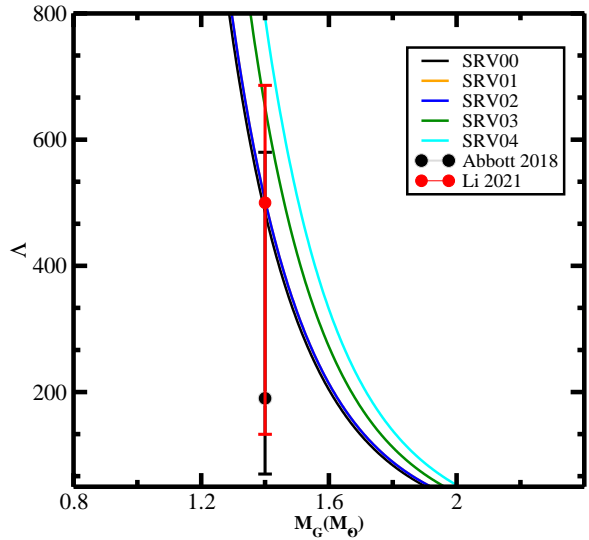}
\caption{\label{tidal} (Color online) Variation of  dimensionless tidal deformability ($\Lambda$)
 with respect
to gravitational mass for SRV  parametrizations.}
\end{figure}

Tidal deformability imparted by the companion stars on one another in a binary system
can yield remarkable information on the EoS for neutron star \cite{Hinderer2008,
Hinderer2010}. In Fig. \ref{tidal}, we show the results of dimensionless tidal deformability
$\Lambda$, defined as $\Lambda$= (2/3)$k_{2}$$(R/M_G)^{5}$, where $k_{2}$ is the love
number, as a function of the neutron star mass $M_G$ for different SRV models.
The recent constraints on the tidal deformability  $\Lambda_{1.4}$ of
1.4$M_{\odot}$  neutron star including GW170817 \cite{Abbott2018,Li2021} is also given
in the figure. The value of $\Lambda_{1.4}$ lies in the range 484 - 783 for different
SRV parametrizations, which satisfies the proposed limit are listed  in Refs. \cite{Abbott2018,
Chen2021,Abbott2017,Reed2021}. The value of  $\Lambda_{1.4}$ increases  with the increase
in the value of the coupling $g_{\delta}$ corresponding to the SRV parametrizations as
can be seen from  Fig. \ref{tidal}. All these results are summarized in Table \ref{tab:table4}. The theoretical erros/uncertainities in   neutron star properties for SRV parametrizations are also mentioned in the table. The neutron star properties such as   $M_{max}$, $R_{max}$, $R_{1.4}$, $R_{2.0}$ are relatively well constrained for all SRV parametrizations (at $\leq$ 3 $\%$)  whereas for $\Lambda_{1.4}$, the theoretical uncertainities are found to be $\leq$ 10$\%$.

To this end, we may mention that  the contribution of $\delta$
meson is considered only through its  linear interaction with
nucleons. However, inclusion of the self interaction of $\delta$ mesons and the
mixed interactions with other mesons may alter the symmetry energy at
suprasaturation densities.  This would  further enhance the flexibility
of EOS of dense matter and accordingly the properties of neutron stars
\cite{Zabari2019,Zabari2019_2,Miyatsu2022}.  Inclusion of the higher order
contributions of $\delta$ meson thus enable one to model the properties of
neutron stars in somewhat independent of the properties of finite nuclei. The optimization of effective Lagrangian that  includes different  terms
involving $\delta$ meson field requires  accurate knowledge  of neutron
star properties over a wide range of mass.

\section{Summary and conclusions}\label{conclusion}
The effect of the isovector-scalar field corresponding to  $\delta$-meson in
relativistic mean field theory is investigated. We have generated five
sets of SRV parametrizations SRV00, SRV01, SRV02, SRV03 and SRV04 to explore
the effects of $\delta$-meson on the properties of finite nuclei, infinite  nuclear
matter, and neutron stars. A covariance analysis to measure the accuracy of
model predictions is also performed. This also enabled us to carry out a
systematic study of correlations among model parameters and various finite
nuclei and infinite nuclear matter properties of interest. The SRV
parametrizations have been obtained in such a way that  they reproduce the
ground state properties of the finite nuclei and infinite  nuclear matter properties
quite convincingly. In turn, they satisfy  the constraints on mass and radius of the neutron star and
its dimensionless  tidal deformability, $\Lambda$, from recent astrophysical
observations \cite{Li2013,Yue2022,Abbott2018,Li2021,Reed2021}. It is observed
that to fit the properties of finite nuclei and infinite nuclear matter, a
stronger coupling between the $\rho$-meson and nucleons ($g_{\rho}$) is required
in the presence $\delta$-meson field. Furthermore, the $\delta$-meson
significantly affects the radius of canonical neutron star. It is found that
the contributions from  $\delta$-meson is important  and   has some
significant effects on the dense matter EoS. The value of $\Lambda_{1.4}$
for different SRV parametrization is also in line with the constraint obtained
from GW170817 event. It is clear that the isovector splitting of effective
mass of nucleon in the presence of $\delta$ in dense asymmetric matter,
like the scenario present in the core of a neutron star, can be significant.
It remains, however, an open question how to identify in future the  signatures of  isovector effective mass splitting
from  astrophysical observations.
\begin{acknowledgments}
Virender Thakur is highly thankful to  Himachal Pradesh University and
DST-INSPIRE (Govt. of India) for  providing computational facility and financial assistance
(Junior/Senior Research Fellowship under grant number “DST/INSPIRE Fellowship/2017/IF170302”).
 C.M. acknowledges partial support
 from the IN2P3 Master Project “NewMAC”.
\end{acknowledgments}

\end{document}